\documentclass[12pt]{article}
\usepackage[italian]{babel}
\usepackage{natbib}\usepackage{txfonts}\usepackage{balance}
\usepackage{graphicx}
\title{L'Eliosismologia: onde sismiche per studiare l'interno del Sole}

\author{M. P. \,Di Mauro \\
\small{INAF--IAPS,
Istituto di Astrofisica e Planetologia Spaziali,}\\
\small{ Via del Fosso del Cavaliere 100, 00133 Roma, Italy}\\
\small \texttt{maria.dimauro@inaf.it}}
\date{}

\begin{document}
\maketitle

\abstract{
Negli ultimi decenni siamo stati testimoni di una straordinaria rivoluzione
 della 
conoscenza e comprensione della nostra stella grazie alla nascita dell'E\-lio\-si\-smo\-lo\-gia, lo studio delle oscillazioni solari.
Analogamente a ci\`o che accade nella Terra durante i terremoti, anche l'interno del Sole \`e pervaso continuamente da onde sismiche che provocano piccole oscillazioni, ovvero deformazioni della fotosfera.
Le oscillazioni sono la manifestazione di diversi processi che avvengono all'interno della struttura del Sole e le frequenze sismiche dei modi osservati e misurati sulla superficie sono legate direttamente ai parametri fisici degli strati interni attraversati dalle onde sismiche. 

In questo articolo 
verranno illustrate le
caratteristiche generali delle oscillazioni solari e verranno
presentati
 i risultati importanti e i progressi notevoli raggiunti grazie all'Eliosismologia.}

%\maketitle{} 

\section{Introduzione}                                                    

E' noto che l'osservazione degli strati superficiali costituisce il mezzo pi\`u immediato per studiare il Sole e,
fino a circa mezzo secolo fa, 
le uniche conoscenze sulla struttura al di sotto della fotosfera  venivano
dedotte dal cosidetto 'modello solare', cio\'e l'insieme delle equazioni
teoriche che ne descrivono lo stato fisico e chimico e che soddisfano le
condizioni al contorno, cio\`e riproducono i parametri noti con certezza quali et\'a, massa, luminosit\'a e composizione chimica superficiale.

Nel 1962, Leighton e i suoi collaboratori (Leighton et al. 1962) scoprirono, grazie ad 
accurate osservazioni, che le righe d'assorbimento dello spettro
fotosferico  del Sole mostravano spostamenti in lunghezza d'onda con
 periodicit\`a di circa 5 minuti.
Questi spostamenti vennero attribuiti 
alla presenza di moti oscillatori di contrazione ed espansione della fotosfera 
solare. 

\begin{figure}[h]
\begin{center}
\includegraphics[width=.7\textwidth]{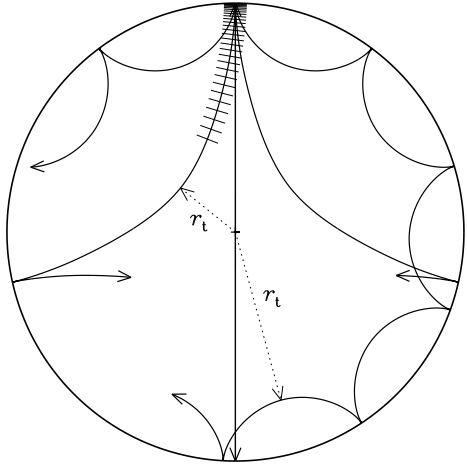}
\caption{Illustrazione schematica della propagazione del suono all'interno
         del Sole. Due modi acustici caratterizzati da diversa lunghezza
         d'onda e frequenza penetrano a profondit\`a $r_t$ differenti.}
\end{center} 
\label{A}
\end{figure}

I piccoli moti della superficie, di ampiezza dell'ordine del decimillesimo di raggio solare,
furono ipotizzati da Ulrich (1970) e Leibacher e Stein (1971)
e poi provati da Deubner (1975)
come dovuti ad onde acustiche, generate nella zona
di convezione e  
intrappolate negli strati interni del Sole come in cavit\`a risonanti 
(vedi Fig.\ref{A}),
 anche se non esistono pareti di riflessione solide.
Un'onda acustica, che si propaga dalla
superficie verso l'interno del Sole (Fig.~\ref{A}), viaggia su una traiettoria a forma di
arco perch\`e, a causa della temperatura crescente verso l'interno,
viene deviata gradualmente fino a
ritornare nuovamente verso la superficie.
L'onda 
intrappolata in uno strato tra la superficie e un'immaginaria parete all'interno, 
a causa di riflessioni successive, interferisce con se stessa dando luogo a configurazioni stazionarie, detti 'modi di oscillazione', ciascuna identificabile con una lunghezza 
d'onda orizzontale 
sulla superficie e con una frequenza. Poich\`e le onde acustiche sono onde associate alla compressione e 
rarefazione del mezzo di propagazione ovvero mantenute da forze di pressione, si \`e soliti parlare di 'modi p'.

\begin{figure}[h]
\begin{center}
\includegraphics[width=.8\textwidth]{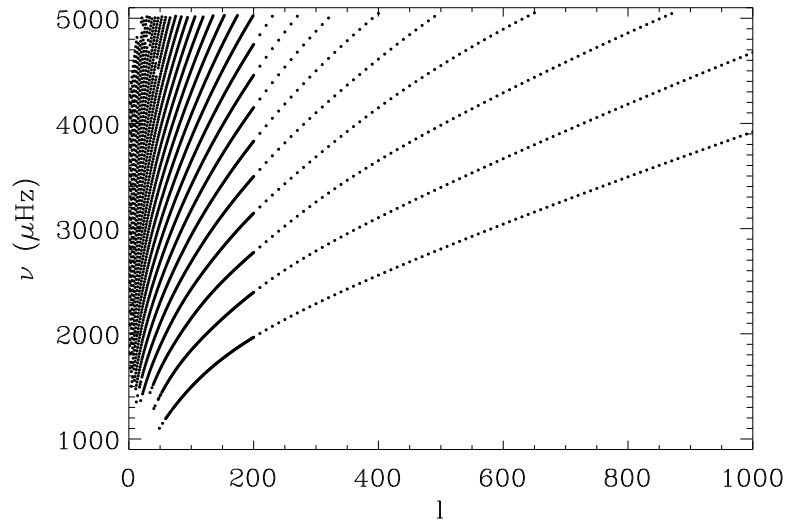}
\caption{Spettro delle oscillazioni solari che mostra il numero dei nodi
  dell'onda sulla superficie l in funzione della frequenza $\nu$ dei modi.}
\label{B}
\end{center}
\end{figure}

Lo spettro di potenza di alcuni modi p di oscillazione del Sole \`e illustrato in Fig.~\ref{B}. Questa figura mostra che le frequenze delle oscillazioni non sono distribuite casualmente, ma ad ogni configurazione ondosa superficiale corrispondono solo un numero finito di 
frequenze cio\'e di modi possibili.

\begin{figure}[h]
\begin{center}
\includegraphics[width=.75\textwidth]{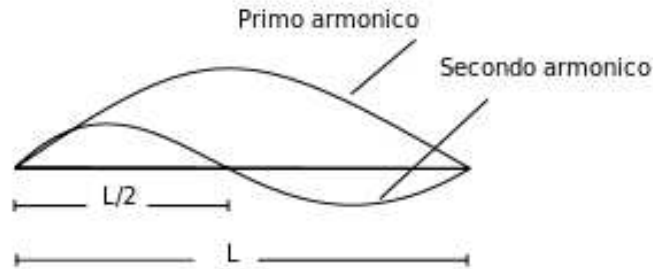}
\caption{Oscillazioni longitudinali di una corda di lunghezza L:
armonica fondamentale  e prima armonica.
 L'armonica fondamentale con due nodi agli estremi e un ventre centrale
 ha frequenza $f=v_s/2L$ e lunghezza d'onda $2L$. La prima armonica con tre nodi di cui due agli estremi e uno centrale e due ventri antisimmetrici, ha frequenza $f_1=2f$ e lunghezza d'onda 
$L$.}
\label{C}
\end{center}
\end{figure}

Per capire meglio cosa succede nel Sole, basta 
pensare ad un esempio pi\'u semplice di corpo in vibrazione: il caso uni-dimensionale di una corda di una chitarra.
Una corda di lunghezza L fissata a due estremi e messa in vibrazione emetter\`a un suono che \`e la combinazione di diverse onde stazionarie, come illustrato in Fig. \ref{C}: l'armonica fondamentale con
 frequenza $f=v_s/2L$ dove $v_s$ \`e la 
velocit\`a del suono, pi\`u le armoniche successive con frequenze che sono multipli interi della frequenza fondamentale,
 cio\'e la prima armonica con frequenza $f_1=2f$, la seconda armonica con frequenza $f_2=3f$ etc...

\begin{figure} 
\includegraphics[width=1.0\linewidth]{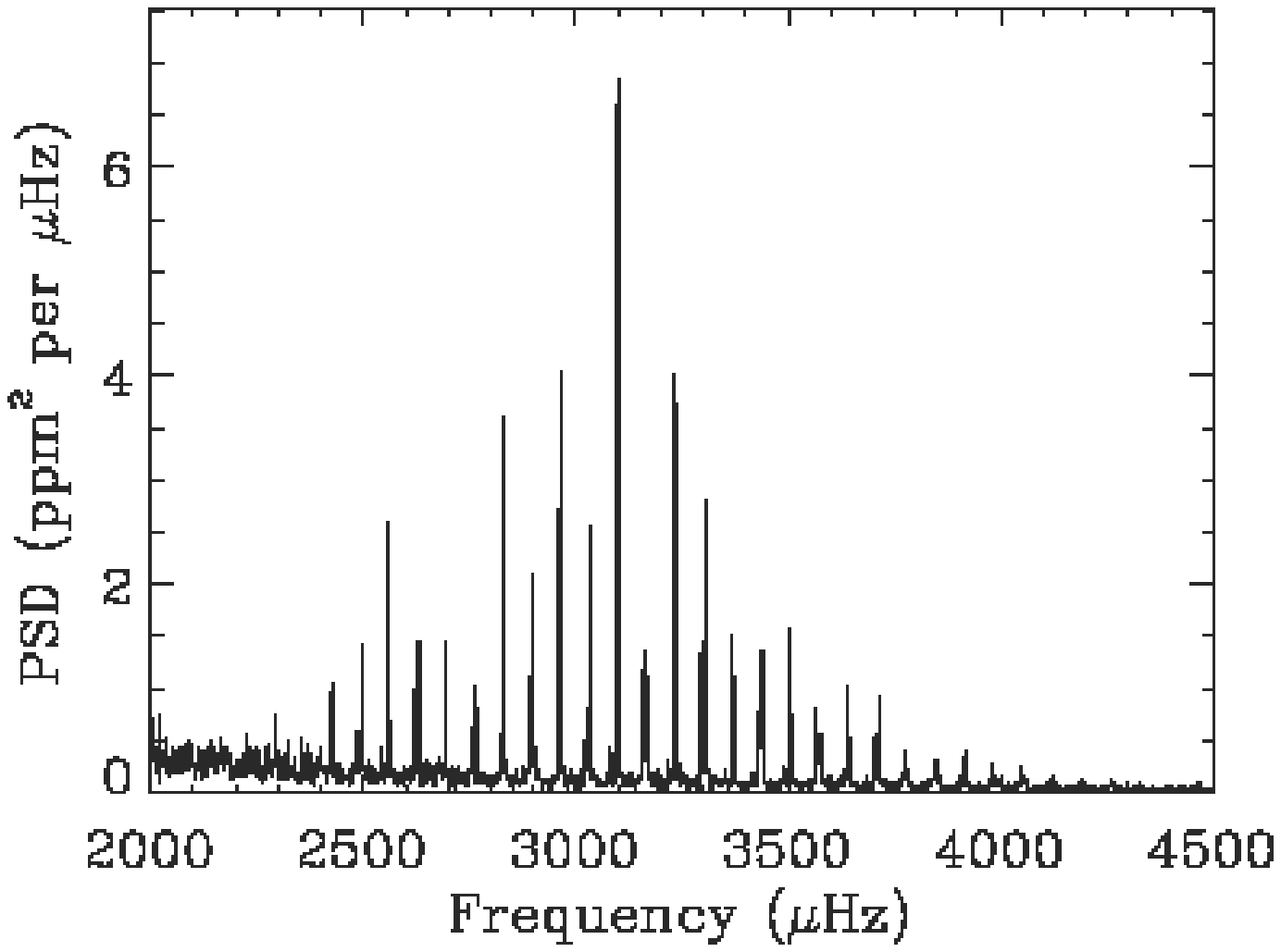} 
\caption{Spettro di oscillazione del Sole.} 
\label{spet} 
\end{figure}

L'aspetto oscillatorio appare complicarsi nel caso bidimensionale, come nei piatti, in cui coesistono diverse configurazioni ondose determinate dalla propagazione di onde circolari e non longitudinali come nella corda.
Nel caso del Sole, le onde che si propagano sono tridimensionali,
e le frequenze mostrate in Fig.~\ref{B} rappresentano per ogni lunghezza
d'onda orizzontale i suoni fondamentali del Sole pi\`u tutti i suoni armonici
con frequenze quasi equispaziate. La Fig.~\ref{spet} mostra lo spettro del Sole,
ovvero le frequenze e le ampiezze dei modi di oscillazione osservati. La
separazione dei modi
nello spettro fornisce informazioni globali sulla massa e sulla et\`a del Sole.

\begin{figure}[htbp]
\begin{center}
\includegraphics[width=.75\textwidth]{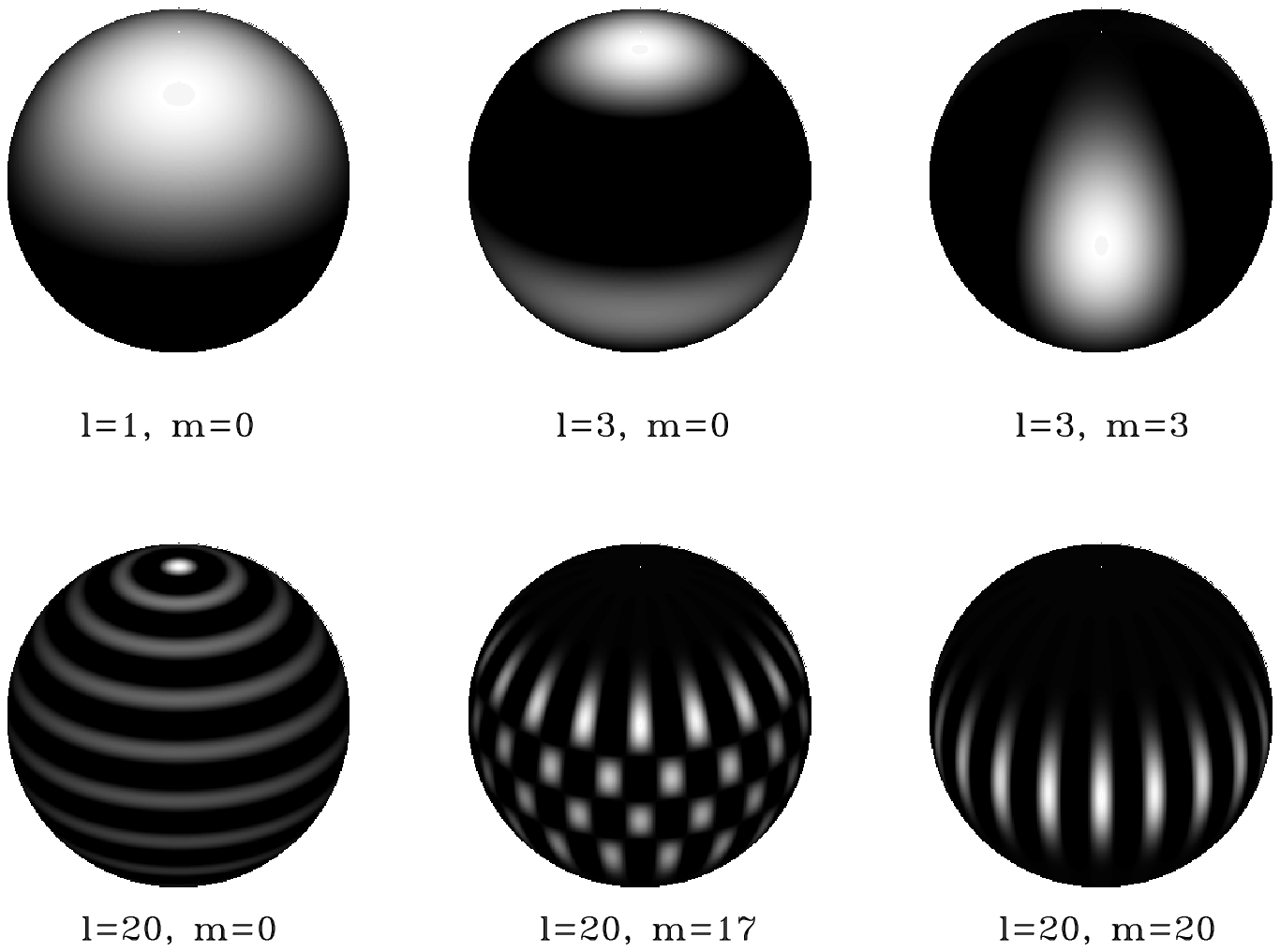}
\caption{Configurazioni spaziali agli occhi di un osservatore, dove l
  rappresenta il numero di nodi lungo la latitudine ed m rappresenta il numero
dei nodi dell'onda lungo la longitudine.}
\label{D}
\end{center}
\end{figure}

Una delle pi\`u importanti caratteristiche delle oscillazioni
del Sole \`e la presenza simultanea
di una moltitudine di modi di oscillazione con frequenze, e quindi periodi, differenti che 
determinano cos\`{\i} una configurazione ondosa estremamente complessa dovuta a milioni di modi 
normali presenti contemporaneamente.
Le piccole oscillazioni
possono essere puramente radiali, nel qual caso il Sole si espande e contrae 
mantenendo la sua forma sferica, o non-radiali, nel qual caso la superficie
solare mostra zone che si espandono simultaneamente a zone contigue che si 
contraggono.\\
La Fig.~\ref{D} mostra alcune configurazioni spaziali di 
 singoli modi in un dato istante: regioni in avvicinamento (in bianco)
alternate a regioni che si allontanano dall'osservatore (in nero).

%All'interno di una stella agiscono principalmente due tipi di forze
%capaci di mantenere le oscillazioni: forza di
%pressione del gas e forza di gravit\`a. Il predominare di una delle due
%forze sull'altra determina rispettivamente onde acustiche e onde  
%di gravit\`a e quindi due differenti nature dei modi di oscillazione. 
%I modi normali relativi ai due tipi di onde
 %prendono il nome di:
%\begin{description}
%\item{a)} {\em modi p}, che sono onde acustiche associate alla compressione e 
%rarefazione del mezzo di propagazione;
%\item{b)} {\em modi g}, che sono onde di gravit\`a prodotte dalla variazione 
%della densit\`a con la profondit\`a e si manifestano allorch\`e la variazione
%di densit\`a tra un elemento di materia in moto e le zone circostanti provoca 
% moti ascendenti e discendenti dovuti al 
%prevalere, o meno,  della forza di gravit\`a sulla spinta di Archimede.
%\end{description}

\section{Eliosismologia}

Il pi\`u importante aspetto delle oscillazioni solari \`e rappresentato, senza
dubbio, dall'approccio sismologico, tramite cui \`e possibile studiare non solo
 la struttura degli strati pi\`u interni, ma anche la
rotazione interna. 
Come ascoltando una melodia \'e possibile
risalire allo strumento musicale che suona, cos\`i dai modi di oscillazione \`e possibile 
ricostruire la struttura, la forma e le caratteristiche del Sole.\\
Sia chiaro per\`o che nessun suono pu\`o giungere dal Sole alla Terra, essendo
troppo bassa la densit\'a delle particelle presenti nello spazio per
permettere una propagazione di onde acustiche, 
ma possiamo facilmente misurare le deformazioni prodotte dai suoni sulla superficie del Sole. 

Le oscillazioni sono la manifestazione dei processi fisici che avvengono 
all'interno del Sole, quindi
 le frequenze osservate contengono un insieme di informazioni sulle
quantit\`a (pressione, densit\`a e loro variazioni) 
che definiscono la struttura delle regioni di propagazione.
In particolare, ogni modo di oscillazione caratterizzato da una frequenza (o periodo) e da
una lunghezza d'onda in superficie penetra e viene intrappolato a 
profondit\`a e a latitudini differenti, per cui sonda una precisa zona 
dell'intera struttura solare.
%La sensibilit\`a delle frequenze alla struttura interna dipende 
%dall'ampiezza delle onde nelle diverse regioni interne.
% L'ampiezza dei {\em modi p} \`e
%piccola vicino al centro per cui anche se 
%possono penetrare vicino al nucleo non possono darne informazioni dettagliate.

I risultati eliosismici pi\'u significativi sono stati ottenuti grazie
 alle misure degli esperimenti GONG (Harvey et al. 1996), IRIS (Fossat et al. 1991), BISON
 (Chaplin et al. 1996 ),
reti di osservatori posti a differenti longitudini che hanno permesso 
l'osservazione ininterrotta del Sole anche per diversi anni. Ma i 
progressi nella determinazione della struttura e
rotazione interna sono stati ottenuti dopo il 1996 in seguito al lancio della
sonda spaziale SOHO (collaborazione NASA ed ESA), i cui due strumenti dedicati alla Eliosismologia,
MDI (Scherrer et al. 1995) e GOLF (Gabriel et al. 1995), hanno prodotto dati con una precisione in frequenza mai raggiunta prima e
 l'identificazione di circa 4000 modi. 
 
\subsection{La struttura interna del Sole}

\begin{figure}[h]
\begin{center}
\includegraphics[width=10cm]{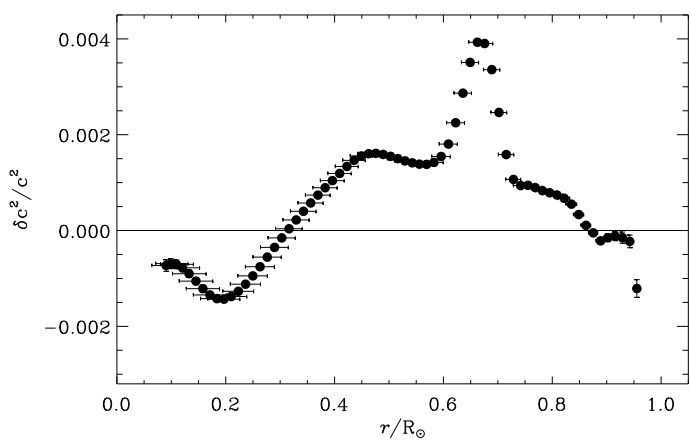}
\caption{Differenza in velocit\`a del suono tra il Sole e il modello 
solare standard.}
\end{center}
\label{Fc2}
\end{figure}

Uno dei risultati fondamentali dell'Eliosismologia \`e l'aver accertato la sostanziale correttezza del 'modello solare standard', termine con cui si definisce l'insieme delle equazioni che descrivono lo stato fisico della nostra stella. 
Le oscillazioni ci hanno rivelato  diversi dettagli strutturali: la posizione della base della zona convettiva, la misura dell'abbondanza di alcuni elementi costituenti, la presenza di fenomeni relativistici nel nucleo e altro ancora.
Il modello solare \'e stato cos\`i rettificato e modificato negli anni e oggi riesce a
riprodurre la struttura interna del Sole con un errore significativamente piccolo (vedi Fig.~\ref{Fc2}). Restano per\`o
ancora delle discrepanze tra il modello e il Sole osservato, in particolare nella zona di
transizione tra la zona radiativa e quella convettiva a circa $0.7$ del
raggio solare e negli strati sotto la
superficie. Questo risultato conferma che non siamo ancora in grado di descrivere le propriet\`a termodinamiche
della zona convettiva e degli strati superficiali.

La comprensione della struttura del Sole \`e stata fondamentale 
per la soluzione di uno dei
 problemi pi\`u discussi della Fisica della particelle e noto come
mistero dei 'neutrini mancanti'. L'Eliosismologia ha contribuito a confermare
la teoria avanzata da Pontecorvo nel 1969 secondo cui
  i neutrini, particelle
che si generano durante la fusione di idrogeno nel nucleo e ipotizzati esistere privi di massa possiedono invece una massa, anche se piccolissima.
Nel lungo viaggio dal Sole alla Terra i neutrini possono cambiare
tipologia. Poich\`e gli esperimenti degli anni '90 erano in grado di rivelare solo neutrini di un tipo (quelli associati agli elettroni), i neutrini che arrivavano a terra come neutrini di un'altra famiglia non venivano contati: in questo modo il flusso osservato risultava inferiore a quello previsto. 
\subsection{La dinamica interna del Sole}

\begin{figure}
 \begin{center}
\includegraphics[width=10cm]{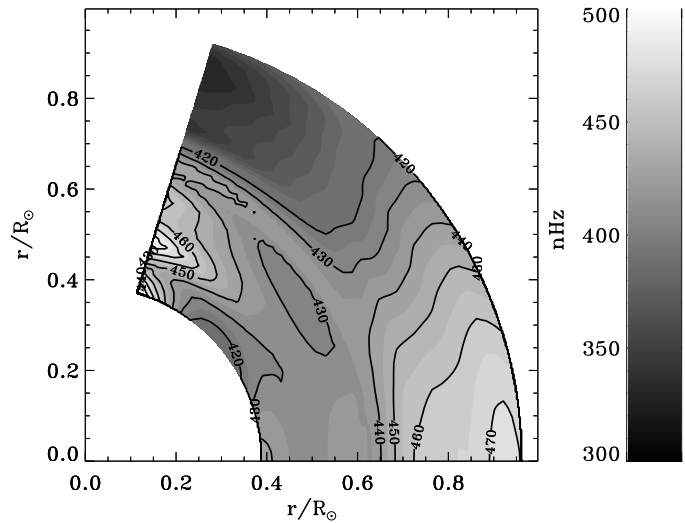}
\caption{Rotazione interna del sole ottenuta
 dai dati ottenuti dallo spazio dallo strumento MDI/SOHO (Schou et
 al. 1998). Colori e contorni indicano zone che ruotano alla stessa velocit\'a.
Le zone bianche sono le regioni per le quali i dati non forniscono
informazioni apprezzabili (Di Mauro et al. 1998).}
\end{center}
\label{rot}
\end{figure}

Per quanto riguarda la dinamica interna,
i risultati eliosimici sono sintetizzati nella Fig. \ref{rot} che mostra una
sezione dell'interno solare. In questa figura i contorni e le diverse gradazioni di grigio
indicano regioni che ruotano a velocit\`a differenti.

I risultati mostrano che la rotazione differenziale osservata in fotosfera, 
pi\`u veloce all'equatore e pi\`u lenta verso i poli,
persiste anche all'interno della zona convettiva, mentre la zona radiativa appare ruotare
come un corpo rigido ad una velocit\`a di circa $430\,{\rm nHz}$ che
corrisponde ad un periodo di circa 27 giorni (Di Mauro et al. 1998, Schou et
al. 1998).
La velocit\`a di rotazione a grandi latitudini aumenta dalla superficie verso l'interno
, mentre a basse latitudini appare decrescere verso l'interno.  
Inoltre la velocit\`a di rotazione non assume valori costanti nel tempo, ma
pare variare con il ciclo di attivit\`a magnetica.

L'Eliosismologia \`e stata in grado di definire, inoltre, le propriet\`a
della cosidetta 'tachocline', regione di transizione dalla rotazione differenziale alla
rotazione rigida (Spiegel et al. ) posizionata a circa $0.71$ del raggio solare.
 Si pensa che l'azione della 'dinamo solare', da cui trae origine il ciclo di
attivit\`a magnetica di $22\,{\rm anni}$, sia indotto dal forte campo
magnetico toroidale creato dalla variazione di rotazione in questo
sottilissimo strato.

Una delle questioni pi\`u importanti e ancora non risolte della dinamica solare \`e la
rotazione del nucleo. I modi p sono poco sensibili
alla struttura del nucleo. Infatti, le onde acustiche alle temperature di milioni di gradi raggiungono velocit\`a cos\`i elevate da sondare per un tempo troppo breve il nucleo.
Per questo motivo, da anni gli scienziati sono alla ricerca dei cosidetti
{\em modi g}, 
 che sono onde di gravit\`a che dovrebbero formarsi nelle zone 
pi\`u interne del Sole a causa
dalla forte variazione 
di densit\`a con la profondit\`a, provocando
 moti ascendenti e discendenti dovuti al 
prevalere, o meno,  della forza di gravit\`a sulla spinta di Archimede.
A differenza dei modi p, i modi g nel Sole sono evanescenti in superficie, ma
hanno 
un'ampiezza
molto grande all'interno, per cui potrebbero dare informazioni estremamente 
dettagliate sulla 
struttura e dinamica del nucleo, qualora venissero identificati.
Solo recentemente, dopo 10 anni di osservazioni
dello strumento GOLF/SOHO \`e stata finalmente annunciata 
l'identificazione di alcuni modi g (Garcia et al. 2007)
La presenza, anche se molto discussa, di questi modi g sembra indicare che il
nucleo solare possa ruotare ad una velocit\`a da 5 a 7 volte pi\`u alta di
quella osservata in
superficie (Garcia et al. 2011).
Questi risultati necessitano per\`o di essere verificati e confermati con
misure ottenute da altri strumenti.

\section{Il futuro dell'Eliosismologia e conclusioni}

Nei capitoli precedenti abbiamo visto che l'Eliosismologia si \`e rivelata
 un chiaro mezzo diagnostico delle condizioni interne del Sole ma,
 malgrado i risultati notevoli e
inaspettati, rimangono ancora molte questioni aperte 
nella comprensione dei fenomeni 
interni del Sole. Per questo motivo,
in questi ultimi anni, diverse
 missioni spaziali sono state progettate allo scopo di ottenere dati eliosismici
ad altissima risoluzione sia spaziale che in frequenza:
SDO, Solar Dynamics Observatory (Schou et al. 2012), una missione della NASA per comprendere meglio l'origine e l'evoluzione del campo magnetico e dei processi dinamici e il loro impatto sulla Terra (lanciato nel 2008);
PICARD (Thuillier et al. 2003), una missione francese del CNES con lo scopo di studiare il clima sulla Terra e le relazioni con la variabili\`a del Sole
 (lanciato nel 2009); e infine
SO, Solar Orbiter (Marsden \& Fleck 2007), un satellite ESA per studiare le regioni vicino ai poli e la faccia del Sole non visibile dalla Terra (lancio previsto nel 2014).

I grandi successi ottenuti dall'Eliosismologia 
hanno spinto recentemente gli scienziati ad estendere questa tecnica anche ad altre stelle
che, come il Sole, mostrano la presenza di deformazioni dovute alle piccole
oscillazioni.  Infatti, modi acustici e modi di gravit\`a possono essere 
generati nelle stelle non solo dai moti convettivi, ma anche attraverso altri meccanismi.
  
Lo studio sismologico delle stelle pulsanti, noto con il nome di 
{\it Astrosismologia}, \`e stato per anni limitato dal problema 
dell'identificazione dei modi.
Infatti, le ampiezze delle oscillazioni
sono molto piccole da poter essere misurate in stelle distanti con
strumenti posti sulla Terra.
Solo recentemente diverse missioni spaziali, tra cui il satellite canadese 
  MOST (Walker et al., 2003), la
  missione francese COROT (Appourchaux et al., 2008), e recentemente la
  missione della NASA
{\it Kepler} (Borucki et al. 2010), progettate e lanciate con successo per la
misura di oscillazioni nelle stelle, hanno svelato 
 risultati impensabili fino a
qualche anno fa.
Le osservazioni ottenute da queste missioni spaziali 
hanno dimostrato che  le oscillazioni sono presenti in stelle di ogni tipo
spettrale e in tutte le fasi evolutive.
 In particolare, 
 oscillazioni acustiche originate, come nel Sole, dai moti convettivi
 sotto la superficie e perci\`o  dette oscillazioni di 'tipo solare',
sono state misurate in un vasto campione di stelle nella fase di 
sequenza principale, nel cui nucleo brucia idrogeno in elio, e anche 
nella fase successiva di sub-gigante, bruciano l'idrogeno in uno strato
sottile attorno al nucleo di elio.
La misura delle separazioni delle frequenze dei modi,
come spiegato nel Capitolo 1,
permette con buona approssimazione la determinazione di
 massa ed et\`a di stelle con oscillazioni di 'tipo solare', 
e anche solo poche frequenze individuate per i modi
 che producono le deformazioni pi\`u evidenti possono fornire
 qualche informazione della struttura interna.

In stelle pi\`u evolute del Sole, che si trovano nella fase di
gigante rossa e che hanno gi\`a o stanno per innescare il bruciamento
dell'elio, sono stati invece inviduati i modi g (Bedding et al. 2010). 
La misura dei modi g, che
sondano le zone del nucleo, ha permesso di determinare con precisione l'et\`a
delle stelle pi\`u evolute.

Appare chiaro come l'Astrosismologia sia fondamentale nella 
comprensione dell'Universo che ci circonda e
in particolare nella ricerca di pianeti simili alla
Terra e nella caratterizzazione di 
stelle e sistemi stellari che ospitano
pianeti extrasolari.
I dati raccolti dalle missioni spaziali per un campione di pi\`u 
di 150.000 stelle
risultano davvero interessanti, ma necessitano ancora di essere tutti
analizzati e interpretati.

Per concludere, i risultati asterosismologici ottenuti grazie alle
osservazioni
delle missioni spaziali, hanno posto l'Eliosismologia e il Sole  
in un contesto pi\`u vasto: tecniche e metodi sviluppati in
Eliosismologia possono essere applicati con successo alle altre stelle, dando
la possibilit\`a di confermare e provare le teorie della evoluzione stellare
e comprendere cos\`i le fasi che portano una stella dalla sequenza principale
 fino alla fase pi\`u evoluta di nana bianca.
L'asteroseismologia fornisce la possibilit\`a di testare e capire
i processi fisici che avvengono nelle stelle. Appare evidente che i
miglioramenti nella caratterizzazione dei modelli stellari
che ne conseguiranno
 saranno cruciali per diversi campi dell'astrofisica, incluso lo studio
 della struttura ed evoluzione della nostra Galassia e la formazione
 degli elementi nell'Universo.

\bibliographystyle{aa}

\newpage
\listoffigures
\end{document}